\documentstyle[aps,epsfig,amssymb,preprint,eqsecnum,floats,cite]{revtex}
\tighten
\def \be{\begin{equation}}
\def \ee{\end{equation}}
\def \bea{\begin{eqnarray}}
\def \eea{\end{eqnarray}}
\def \ben{\begin{enumerate}}
\def \een{\end{enumerate}}
\def \bit{\begin{itemize}}
\def \eit{\end{itemize}}
\def \baR{\begin{array}}
\def \eaR{\end{array}}
\def \GeV{{\text{GeV}}}

\def \MeV{{\text{MeV}}}
\def \TeV{{\text{TeV}}}

\def \bm{\boldmath}
\def \B{\bar{B}}

\def \Bd{\bar{B}_d \to \mu^+\mu^-}
\def \Bs{\bar{B}_s \to \mu^+\mu^-}
\def \Bq{\bar{B}_q \to  l^+ l^-}
\def \Bds{\bar{B}_{d,s} \to  \m^+ \m^-}
\def \br{{\cal B}\,}
\def \cl#1{{#1\%\ \rm{C.L.}}}

\def \del#1{\Delta^{#1}_{\tilde D_L}}
\def \delu#1{\Delta^{#1}_{\tilde U_L}}
\def \eq#1{Eq.~(\ref{#1})}
\def \eqs#1#2{Eqs.~(\ref{#1}) and (\ref{#2})}
\def \f{\frac}
\def \fig#1{Fig.~\ref{#1}}

\def \glu{\tilde{g}}
\def \hc{\mathrm{H.c.}}
\def \ml#1{M^2_{\tilde{#1}_{L}}}
\def \mlii#1{(M_{\tilde{#1}_{L}})_{ii}}
\def \mrii#1{(M_{\tilde{#1}_{R}})_{ii}}
\def \M{{\cal M}}
\def \mmuhat{\hat m_\m}
\def \neu{\tilde{\chi}^0}
\def \op{{\cal O}}

\def \rf{Ref.~\cite}
\def \rfs{Refs.~\cite}
\def \sec#1{Sec.~\ref{#1}}
\def \sdown{\tilde{d}}
\def \sel{\tilde{l}}
\def \sup{\tilde{u}}
\def \ti{\widetilde}
\def \unit{\leavevmde\hbox{\small1\kern-3.6pt\normalsize1}}
\def \vckm{V_{\text{CKM}}}

\def \chargino{\tilde{\chi}^{\pm}}

\def \cl#1{{#1\%\ \mathrm{C.L.}}}

\def \Im{{\text{Im}}\,}

\def \bm{\boldmath}

\def \bracket#1#2#3{\langle #1|#2| #3\rangle}

\def \cp{\mathrm{CP}}

\def \diag{{\mathrm{diag}}}

\def \ea{{\it et al.}}

\def \eq#1{Eq.~(\ref{#1})}
\def \eqs#1#2{Eqs.~(\ref{#1})--(\ref{#2})}
\def \fig#1{Fig.~\ref{#1}}

\def \nnu{\nonumber}

\def \Oi{{\mathcal O}}

\def \sm{\mathrm{SM}}
\def \rf{Ref.~\cite}
\def \rfs{Refs.~\cite}
\def \sec#1{Sec.~\ref{#1}}

\def \a{\alpha}
\def \b{\beta}
\def \D{\Delta}
\def \g{\gamma}
\def \G{\Gamma}
\def \d{\delta}

\def \m{\mu}
\def \n{\nu}

\def \p{\pi}

\def \s{\sigma}

\def \t{\tau}
\def\euro#1#2#3{{Eur. Phys. J. C} {\bf #1}, #3 (#2)}

\def\ibid#1#2#3{{\it ibid.\/}~{\bf#1}, #3 (#2)}
\def\jhep#1#2#3{{J.~High~Energy~Phys.}~{\bf #1}, #3 (#2)}

\def\nim#1#2#3{{Nucl.~Instrum.~Meth. A}~{\bf #1}, #3 (#2)}
\def\np#1#2#3{{Nucl.~Phys.}~{\bf B#1}, #3 (#2)}
\def\npps#1#2#3{{Nucl.~Phys.~B  (Proc.~Suppl.)}~{\bf #1}, #3 (#2)}
\def\pl#1#2#3{{Phys.~Lett. B}~{\bf #1}, #3 (#2)}

\def\prd#1#2#3{{Phys.~Rev. D}~{\bf #1}, #3 (#2)}
\def\prl#1#2#3{{Phys.~Rev.~Lett.}~{\bf #1}, #3 (#2)}
\def\prp#1#2#3{{Phys.~Rep.}~{\bf #1}, #3 (#2)}

\begin{document}
\preprint{\setlength{\baselineskip}{1.5em}
\small
\vbox{\vspace{-4cm}
\hbox{TUM-HEP-458/02}
\hbox{hep-ph/0204225}
\hbox{April 2002}}}
\draft
\title{\bm Enhancement of $\br(\Bd)/\br(\Bs)$ in the MSSM  with
  Modified Minimal Flavour Violation and Large $\tan\b$}
\author{\sc C.~Bobeth, T.~Ewerth, F.~Kr\"uger, and 
J.~Urban\thanks{\footnotesize E-mail addresses: 
bobeth@ph.tum.de, tewerth@ph.tum.de, fkrueger@ph.tum.de, urban@ph.tum.de}}
\address{Physik Department, Technische Universit\"at M\"unchen,
  D-85748  Garching, Germany}
\maketitle
\begin{abstract}
We extend our previous analysis of the decays $\Bds$ in the Mini\-mal
Supersymmetric  Standard Model (MSSM) to include gluino and neutralino
contributions.  We provide analytic formulae, valid at large values of 
$\tan\b$, for the scalar and pseudoscalar Wilson coefficients arising
from neutral  Higgs boson exchange diagrams with gluinos and
neutralinos.~Together  with the remaining contributions ($W^\pm$,
$H^\pm$, $\chargino$),  and assuming the
Cabib\-bo-Ko\-ba\-ya\-shi-Maskawa (CKM) matrix to be the only source
of flavour  violation, we assess
their implications for the branching fractions $\br(\Bds)$. Of 
particular interest is the quantity $R\equiv\br(\Bd)/\br(\Bs)$, since
(i) the  theoretical errors cancel to a large extent, and (ii) it
offers a  theoretically clean way of extracting the ratio
$|V_{td}/V_{ts}|$  in the Standard Model, which predicts
$R_{\sm}\sim |V_{td}/V_{ts}|^2\sim
O(10^{-2})$. Exploring three different scenarios of modified 
minimal flavour violation ($\overline{\rm MFV}$), 
we find that part of the MSSM parameter space can accommodate  large $\Bds$ 
branching fractions, while being consistent with various experimental
constraints.  More importantly, we show that the ratio $R$ can be as
large as $O(1)$, while the individual branching fractions may be
amenable  to detection by ongoing experiments. We conclude that within 
the MSSM with large $\tan\b$ the decay rates of $\Bd$ and $\Bs $ can
be of comparable size even in the case where 
flavour violation is due solely to the CKM matrix.
\end{abstract}


\section{Introduction}
In the Standard Model (SM), the dominant contributions to the decays
$\Bq$,  where $q=d,s$ and $l=e,\m,\t$, arise from $Z^0$-penguin
diagrams and box  diagrams involving $W^\pm$ bosons.~The contributions
due to  neutral Higgs boson exchange, on the other hand, are
suppressed by a  factor of $m_l m_{b,q}/M_W^2\lesssim 10^{-3}$, and
therefore are completely negligible.

The SM predicts the $\Bs$ branching ratio to be \cite{lectures,cscp} 
\be\label{sm:Bs}
 \br(\Bs)=(3.2\pm1.5)\times 10^{-9},
\ee
and the ratio of  branching  fractions
\be\label{rsm}
 R_{\rm SM} \equiv \left. \f{\br(\Bd)}{\br(\Bs)} \right|_{\rm SM}
   \approx \f{\tau_{B_d}}{\tau_{B_s}}\f{M_{B_d}}{M_{B_s}}
   \f{f_{B_d}^2}{f_{B_s}^2} \f{|V_{td}|^2}{|V_{ts}|^2} \sim O(10^{-2}),
\ee
where $\tau_{B_q}$ is the lifetime of the $B_q$ meson, $M_{B_q}$ and
$f_{B_q}$ are the corresponding mass and decay constant.~We note that
the main uncertainty on the branching ratio in \eq{sm:Bs} arises from
$f_{B_s}$, with a typical value of $(230\pm 30)\ \MeV$ obtained from
lattice QCD calculations \cite{lattice}. On the other hand, the
relative rates of $\B_d$ and $\B_s$ decays [\eq{rsm}] have a smaller
theoretical uncertainty due to the appearance of the ratio
$f_{B_d}/f_{B_s}$,  which can be more precisely determined 
than $f_{B_s}$ alone. A determination of $R$ can therefore potentially
provide valuable information on $|V_{td}/V_{ts}|$ in the
SM. However, given the SM prediction of $\br(\Bd)\sim O(10^{-10})$,
the $\Bd$ decay is experimentally remote unless it is significantly
enhanced  by new physics.~Thus, the purely leptonic decays of 
neutral $B$ mesons provide an ideal testing ground for physics outside
the  SM, with the current experimental upper bounds \cite{exp:bmumu} 
\begin{mathletters}\label{Bmumu:exp:bounds}
\be
 \br(\Bd) < 2.8 \times 10^{-7} \quad (\cl{90}),
\ee
\be
 \br(\Bs) < 2.0 \times 10^{-6} \quad (\cl{90}).
\ee
\end{mathletters}%

Our main interest in this paper is in a  qualitative comparison  of
the  $\Bds$ branching fractions in the presence of non-standard
interactions, which can be made by using the ratio
\be\label{ratio:BdBs}
 R\equiv \frac{\br(\Bd)}{\br(\Bs)}.
\ee
Referring to \eq{rsm}, it is important to note that the suppression of
$R$ in the SM  is largely due to the ratio of the
Cabib\-bo-Ko\-ba\-ya\-shi-Maskawa  (CKM) elements.~This dependence
on the CKM factors  allegedly pertains to all models in which the quark 
mixing matrix
is the  only source of flavour violation. It is therefore interesting
to  ask if $R$ could be of the order unity in some non-standard 
models where flavour violation is governed exclusively by the CKM matrix.
Working in the framework of the Minimal
Supersymmetric  Standard Model (MSSM) with a large ratio of Higgs
vacuum  expectation values, $\tan\b$  (ranging from $40$ to $60$),
we show that such a scenario does exist, and study its consequences for 
the $\Bds$ branching ratios.\footnote{A number of authors \cite{cscp,Bll:SUSY,chan:slaw,huang:etal} have
noted that in  the large $\tan\b$ regime 
neutral Higgs-boson contributions can increase the SM
decay  rates of 
$\Bds$ by up to several orders of magnitude. Moreover, there 
is a correlation between $\br(\Bs)$ and $B_s^0$--$\bar{B^0_s}$ mixing
\cite{oszi,scen:B}, and, in mSUGRA scenarios, between 
$\br(\Bs)$ and $(g-2)_\mu$  \cite{Bmumu:mSUGRA}
(see also Ref. \cite{Bmumu:mSUGRA2}).} To this end, we extend our previous
analysis  \cite{cscp} to include the effects of gluino and neutralino
contributions,  in addition to those coming from diagrams with
$W^\pm, \chargino$, charged and neutral Higgs boson exchange.

The outline of this paper is as follows.~In \sec{mssm}, we 
define modified minimal flavour violation ($\overline{\rm MFV}$) and
discuss briefly three  distinct scenarios within 
the MSSM.~The effective Hamiltonian describing the decays $\Bds$ in
the presence of  non-SM interactions is given 
in \sec{heff:branch}, together with the branching ratios. In
\sec{computation},  we present analytic formulae for 
the gluino and neutralino contributions to scalar and pseudoscalar
Wilson coefficients governing the $b \to q l^+l^-$
transition. Numerical  results for the branching fractions $\br(\Bds)$
and the  ratio $R$ are presented in \sec{num:analysis}. In
\sec{discussion},  we summarize and conclude.

\section{The MSSM with Modified Minimal Flavour Violation}\label{mssm}
%
There exists no unique definition  of minimal flavour violation (MFV)
in the literature 
(see, e.g., Refs. \cite{oszi,laplace,uut,london,munich}). 
The common feature of these MFV definitions is that flavour violation
and/or flavour-changing neutral current (FCNC) 
processes are entirely governed by the 
CKM matrix. On the other hand, they differ, for example, by the
following additional assumptions:
\begin{itemize}
\item[i)] there are no new operators present, in addition to those of the SM 
          \cite{lectures, oszi,laplace,{uut}},
\item[ii)] FCNC processes are proportional to the same combination of
  CKM elements as in the SM \cite{london},
\item[iii)] flavour transitions occur only in charged currents at tree
  level \cite{{munich}}.
\end{itemize}
While these ad hoc assumptions are useful for certain considerations,
such as the construction of the universal unitarity triangle \cite{uut}, 
they cannot be justified by symmetry arguments on the level of the 
Lagrangian. For example, the number
of operators with a certain dimension is always fixed by the symmetry of the 
low-energy effective theory.~Whether the
Wilson coefficients are negligible or not, depends crucially  on the
model considered and on the part of the parameter space. Furthermore, the 
requirement that FCNC processes 
are proportional to the same combination of CKM elements as in the SM
fails, for example, in the MSSM and can be retained only after further simplifying assumptions.
The last statement in iii) is of pure phenomenological 
relevance in order to avoid huge contributions to FCNC processes.

Using symmetry arguments, we propose an approach that  relies only on the 
key ingredient of the MFV definitions in 
\rfs{{oszi},{laplace},{london},{munich},{uut}}, without considering the 
above mentioned additional assumptions i)--iii). We call an extension of the SM
a modified minimal flavour-violating ($\overline{\rm MFV}$) model if
and only if FCNC processes or flavour violation are
entirely ruled by the CKM matrix; that is, we require that FCNC
processes 
vanish to all orders in perturbation theory in the limit $\vckm \to \openone$.

Let us briefly motivate our definition of $\overline{\rm MFV}$.
The structure of the SM is such that the conservation of lepton
numbers $L_e$, $L_\mu$ and $L_\tau$ is exact. This is reflected by 
three global $U(1)$ symmetries (one for each lepton flavour) of the SM 
Lagrangian. 
As a consequence, there are no FCNC processes in the lepton sector to all 
orders in perturbation theory. We regard the breaking of these lepton
flavour $U(1)$ symmetries as flavour violation.\footnote{This is
  different from Ref. \cite{def:MFV}, where flavour violation is regarded as the breaking of
the flavour symmetry $G_F$.} As far as the quark sector of the SM is
concerned, the CKM matrix explicitly breaks the quark flavour $U(1)$ 
symmetries.
Thus, in the quark sector, FCNC processes can be generated in
electroweak interactions. In the limit $\vckm \to \openone$
the three $U(1)$ quark symmetries are restored and therefore FCNC 
transitions vanish to all orders in perturbation theory. This observation
leads us to the definition of $\overline{\rm MFV}$ given above.
As will become clear, the advantage of  
$\overline{\rm MFV}$ is that it is less
restrictive than MFV, while the CKM matrix remains the only source of 
FCNC transitions.

Our definition of $\overline{\rm MFV}$ is manifest basis independent. 
However, in order to find a useful classification of different
$\overline{\rm MFV}$ scenarios within the MSSM, we will  work
in the super-CKM basis (see \rf{fcnc}  for details). In this basis
the quark mass matrices are diagonal,
and  both quarks and squarks are rotated simultaneously. The scalar quark 
mass-squared matrices in this basis have the structure 
\be\label{squark:mass}
 \M_U^2 = \left(\begin{array}{cc} \M^2_{U_{LL}} & \M^2_{U_{LR}}\\ 
                \M^{2\dag}_{U_{LR}} & \M^2_{U_{RR}}\end{array}\right), \quad
 \M_D^2 = \left(\begin{array}{cc} \M^2_{D_{LL}} & \M^2_{D_{LR}}\\
                \M^{2\dag}_{D_{LR}} & \M^2_{D_{RR}}\end{array}\right),
\ee
where the $3\times 3$ submatrices are given by 
\begin{mathletters}
\be 
\M^2_{U_{LL}} = \ml{U} + M_U^2  + \frac{1}{6} M_Z^2\cos2\b(3 - 
4 \sin^2\theta_W)\openone, 
\ee 
\be
\M^2_{U_{LR}} = M_U(A_U^\ast-\m\cot\b\openone),
\ee 
\be
\M^2_{U_{RR}} = M_{\tilde U_R}^2 + M_U^2 + 
\frac{2}{3}M_Z^2\cos 2\b\sin^2\theta_W\openone,
\ee
\end{mathletters}
\begin{mathletters}
\be                     
 \M^2_{D_{LL}} = \ml{D}+ M_D^2 - 
\frac{1}{6}M_Z^2\cos2\b(3-2\sin^2\theta_W)\openone, 
\ee
\be
\M^2_{D_{LR}} = M_D (A_D^\ast-\m\tan\b\openone),
\ee 
\be
\M^2_{D_{RR}} = M_{\tilde D_R}^2 + M_D^2 - \frac{1}{3}M_Z^2\cos 2\b
\sin^2\theta_W\openone.
\ee
\end{mathletters}%
Here, $M_{\tilde{U}_{L,R}}^2$ and $M_{\tilde{D}_{L,R}}^2$ are the soft
SUSY  breaking squark mass-squared matrices,  
$\m$ is the Higgsino mixing parameter, $A_U$ and $A_D$ are soft
SUSY  breaking trilinear couplings, $\openone$ denotes the $3\times
3$  unit matrix, and 
\be
M_U\equiv \diag(m_u, m_c,m_t), \quad M_D\equiv \diag(m_d, m_s,m_b).
\ee
Because of SU(2) gauge invariance, the mass matrix $ \ml{D}$ is
intimately  connected to $\ml{U}$ via 
\be\label{su2}
 \ml{D}=\vckm^\dag \ml{U} \vckm,
\ee
which is important for our subsequent discussion. In the case of
$\overline{\rm MFV}$, the matrices $A_U,A_D$, 
$M^2_{\tilde U_{R}},M^2_{\tilde D_{R}}$ must be diagonal: 
\be
 A_U \equiv \diag(A_u,A_c,A_t), \quad A_D \equiv  \diag(A_d,A_s,A_b),
\ee
\be
 M^2_{\tilde U_{R}} \equiv  \diag(m^2_{\tilde u_{R}}, m^2_{\tilde
   c_{R}},  m^2_{\tilde t_{R}}),\quad 
 M^2_{\tilde D_{R}} \equiv  \diag(m^2_{\tilde d_{R}}, m^2_{\tilde
   s_{R}},  m^2_{\tilde b_{R}}).
\ee
Furthermore, assuming that there are no new CP-violating phases, in
addition  to the single CKM phase, the 
matrices $A_U$ and $A_D$, as well as $\m$, are real.

Taking into account the relation in \eq{su2}, one encounters three
cases of $\overline{\rm MFV}$. 
\bit
\item {\bf Scenario (A):}\\
$\ml{U}$ is proportional to the unit matrix, and so
$\ml{U}=\ml{D}$.~As a result, there  are no gluino and neutralino
contributions  to flavour-changing transitions at one-loop level. The
implications  of this scenario for rare $B$ decays have been
discussed,  e.g., in \rfs{cscp,chan:slaw,huang:etal,Bll:SUSY}. 
This scenario of $\overline{\rm MFV}$ coincides with the
MFV scenario at low $\tan\beta$, as defined in Refs. \cite{{munich},{uut}}.  
\item {\bf Scenario (B):}\\
$\ml{D}$ is diagonal but not proportional to the unit matrix and, in 
consequence,
$\ml{U}$ has  non-diagonal entries.~In such a case, there are again no
gluino and neutralino contributions to flavour-changing one-loop
transitions involving only external down-type quarks and
leptons. However, 
additional chargino contributions show up, due to
non-diagonal entries of $\ml{U}$. 
This scenario has recently been investigated in Ref.~\cite{scen:B}.
\item {\bf Scenario (C):}\\
$\ml{U}$ is diagonal but not proportional to the unit matrix, which
gives  rise to off-diagonal entries in 
$\ml{D}$.~Accordingly, gluino and neutralino exchange diagrams
(in addition to those involving $W^\pm, \chargino$,
charged  and neutral Higgs bosons)
contribute to flavour-changing transitions at 
one-loop level that involve external down-type quarks.
\eit

The common feature of  all these scenarios is that the CKM matrix is
the only  source of flavour violation.
For the remainder of this paper we will concentrate mainly on
scenarios (B)  and (C). To our knowledge, 
the consequences of scenario (C) for the decays $\Bds$ have not yet
been  discussed in the literature. 

\section{Effective Hamiltonian and  branching ratio}\label{heff:branch}
The effective Hamiltonian responsible for the processes $\Bq$, with
$q=d,s$  and $l=e,\m,\t$, in the presence of non-standard interactions
is  given by\footnote{We omit the index $q$ on the operators 
and the corresponding short-distance coefficients.}
\be\label{heff}
 H_{\rm eff} = -\f{G_F\a}{\sqrt{2}\p} V_{tb}^{}V_{tq}^{*}
               \sum_{i=10,S,P}[c_i(\m)\op_i(\m)+c_i'(\m)\op_i'(\m)],
\ee
with the short-distance coefficients $c_i^{(\prime)}(\m)$ and the
local  operators
\begin{mathletters}\label{operator:basis}
\be
 {\Oi}_{10}=(\bar{q} \gamma^{\mu} P_L b) (\bar{l} \gamma_{\mu}\gamma_5 l),\quad
 {\Oi}_{10}^\prime = (\bar{q} \gamma^{\mu} P_R b) (\bar{l} \gamma_{\mu}\gamma_5 l),
\ee
\be
 {\Oi}_S= m_b (\bar{q} P_R b) (\bar{l}l),\quad
 {\Oi}'_S= m_q (\bar{q} P_L b) (\bar{l}l),
\ee
\be
 {\Oi}_P= m_b (\bar{q} P_R b)(\bar{l} \gamma_5 l),\quad
 {\Oi}'_P=m_q (\bar{q} P_L b)(\bar{l} \gamma_5 l),
\ee
\end{mathletters}%
where $P_{L,R}= (1\mp \g_5)/2$. In addition to the operators in
Eqs.~(\ref{operator:basis}),  there are tensor 
and vector operators, 
$(\bar{q} \sigma^{\mu\nu} P_{L,R} b)(\bar{l} \sigma_{\mu\nu} P_{L,R} l)$ 
and $(\bar{q} \gamma^{\mu} P_{L,R} b)(\bar{l} \gamma_{\mu}
l)$. However, they  do not contribute to the $\Bq$ 
decays and do not mix with those appearing in \eq{heff}. As a matter
of fact,  the matrix element that involves the 
antisymmetric tensor $\s^{\m\n}$ must vanish since $p^\m\equiv
p_{B_q}^\m$ is  the only four-momentum 
vector available. Similarly, the matrix element 
$\bracket{l^+l^-}{\bar l\g_\m l}{0}$ does not contribute when 
contracted with 
$\bracket{0}{\bar{q} \gamma^{\mu} P_{L,R} b}{\B_q(p)}\propto p^\m $ 
by the equation of motion.

Evolution of the short-distance coefficients $c_i^{(\prime)}(\m)$ from
the  matching scale $\m_t =m_t^{\rm pole}$ 
down to the low-energy scale $\m_b=m_b^{\rm pole}$ can be performed by
means  of the renormalization group equation. We note that the
anomalous  dimensions of the operators in Eqs.~(\ref{operator:basis})
vanish,  and thus the renormalization group evolution is trivial.

Because of the pseudoscalar nature of the decaying $B_q$ meson, the
hadronic  matrix elements are non-zero 
only for axial-vector and pseudoscalar operators; namely,  
\be\label{axial-vector:me}
 \bracket{0}{\bar{q}\g_{\m}\g_5 b}{\B_q(p)}=ip_{\mu}f_{B_q},
\ee
and, employing the equation of motion, 
\be\label{matrix:element:btoll:II}
\bracket{0}{\bar{q}\g_5 b}{\B_q(p)}=- i f_{B_q} \frac{M_{B_q}^2}{m_b+m_q}.
\ee
Here $f_{B_q}$ is the $B_q$ meson decay constant, which can be
obtained from  lattice QCD computations \cite{lattice}:
\be\label{decay-constants}
 f_{B_d}=(200\pm 30)\ \MeV,\quad f_{B_s}=(230\pm 30)\ \MeV, \quad
 \f{f_{B_s}} {f_{B_d}} = 1.16\pm 0.04.
\ee
(Similar results have recently been obtained from a QCD sum rule
analysis \cite{jamin:lange}.)

If the lepton spins are not measured, the branching ratio for the case
$l=\m$  takes the general form 
\be\label{Btomumu:BR}
 \br(\B_q\to \m^+\m^-) = \f{G_F^2 \a^2 M_{B_q}^3 f_{B_q}^2
   \tau_{B_q}}{64  \pi^3}
 |V_{tb}^{}V_{tq}^{\ast}|^2\sqrt{1- 4 \mmuhat^2} 
 \Bigg\{ ( 1- 4\mmuhat^2 ) |F_S|^2+ | F_P+ 2\mmuhat F_A|^2 \Bigg\},
\ee
with the notation $\mmuhat\equiv m_\m/M_{B_q}$ and the dimensionless
form  factors
\be\label{form:fac}
 F_{S,P} = M_{B_q}\Bigg[\f{c_{S,P} m_b-c_{S,P}' m_q}{m_b+m_q} \Bigg],
           \quad F_A = c_{10}-c_{10}^\prime.
\ee
In the SM, the contributions involving the neutral Higgs boson are
completely  negligible, and so $\br(\B_q\to \m^+\m^-)\propto
\mmuhat^2$, which  is a consequence of helicity suppression.\footnote{Because 
the $B$ meson has spin zero, both  $\m^+$ and $\m^-$ must have the
same  helicity.} Finally, allowing the input parameters in
\eq{Btomumu:BR} to  vary over the interval in \eq{decay-constants},
and using the ranges for the CKM factors given in \rf{lectures}, the
ratio of decay rates  of $\B_{d,s} \to \m^+\m^-$ within the SM is 
expected to be in the range
\be
 0.02 \lesssim R_{\sm}  \lesssim 0.05,
\ee
which is largely due to the imprecisely known CKM elements.

\section{Higgs-boson contributions to the decays 
\bm$\lowercase{\B_{d,s}\to l^+ l^-}$}\label{computation}
We now turn to the computation of the scalar and pseudoscalar Wilson
coefficients in the $b\to ql^+l^-$  transition arising from gluino and
neutralino exchange diagrams within the general MSSM. As mentioned earlier, 
we perform our 
calculation in the large $\tan\b$ regime 
(i.e. $40 \leqslant \tan\b \leqslant 60$). For the remaining 
contributions ($W^\pm, H^\pm,\chargino$) to these short-distance
coefficients, we refer to \rfs{cscp,chan:slaw,huang:etal,Bll:SUSY}.     

\begin{figure}[t]
\begin{center}
\epsfig{file=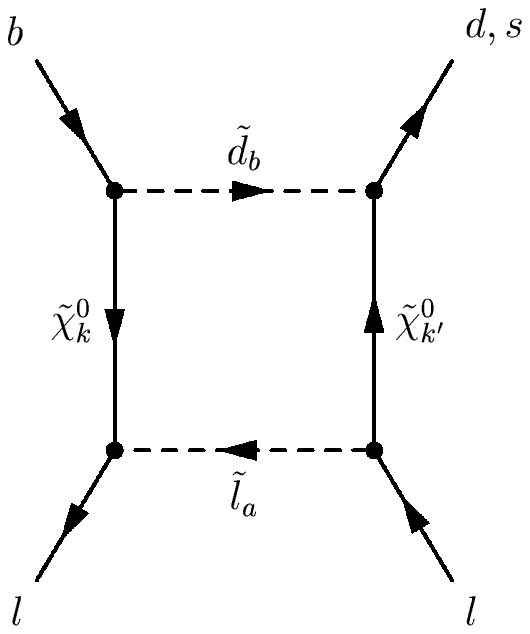,height=1.9in}\hspace{3em}
\epsfig{file=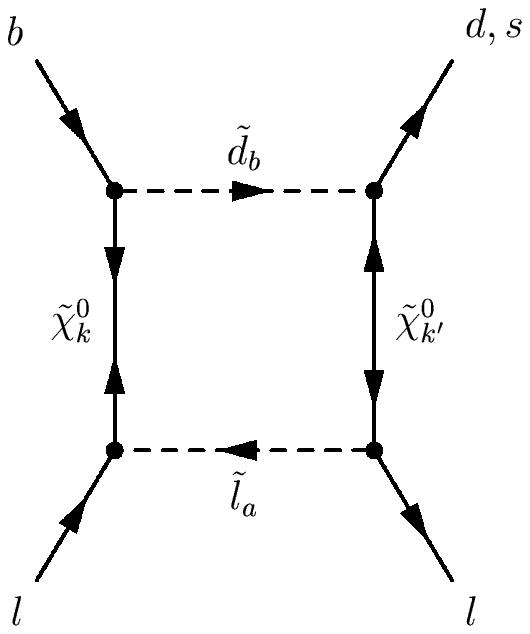,height=1.9in}\hspace{3em}
\epsfig{file=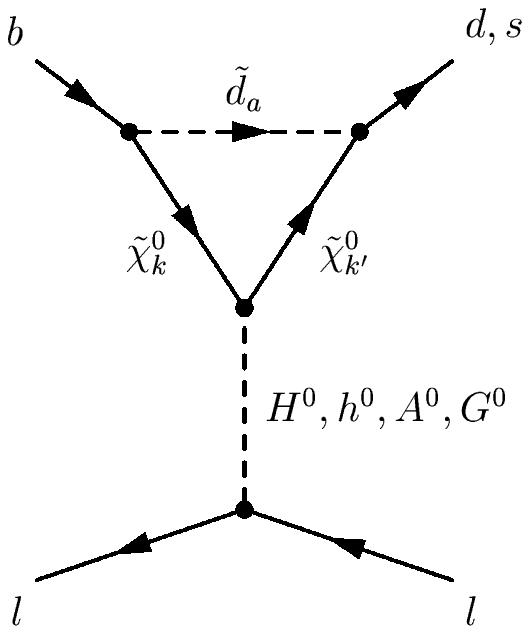,height=1.9in}\vspace{1em}
\epsfig{file=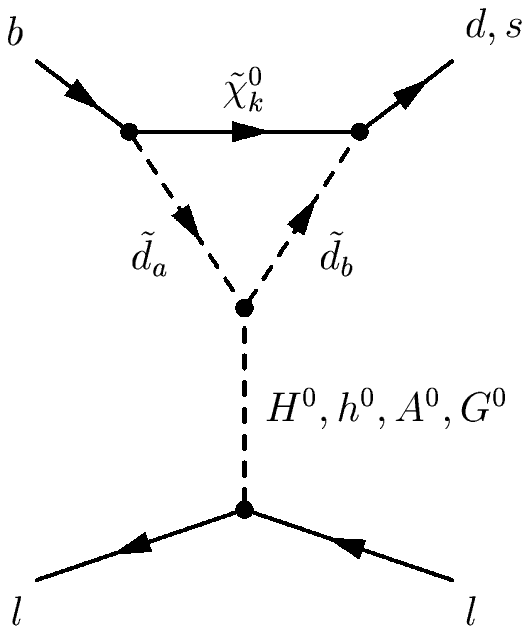,height=1.9in}\hspace{3em}
\epsfig{file=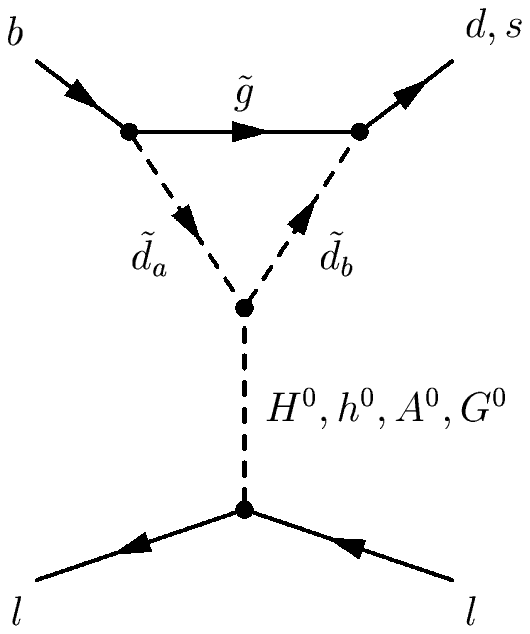,height=1.9in}\hspace{3em}
\epsfig{file=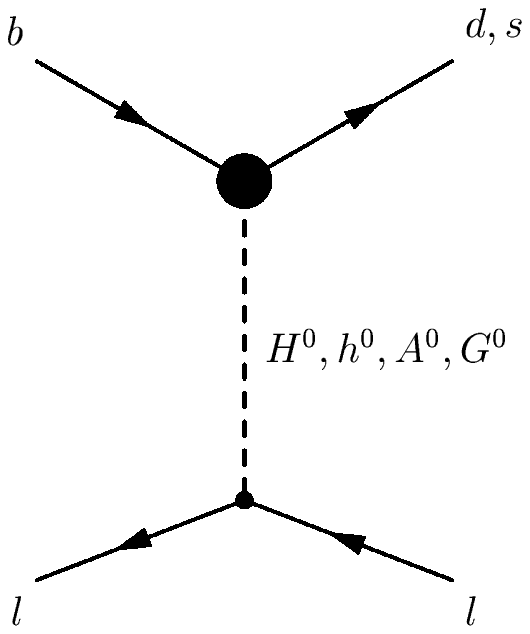,height=1.9in}\vspace{1em}
\caption{Gluino and neutralino diagrams contributing to the decay 
$\B_{d,s}\to l^+l^-$, 
with $l=e,\m,\t$.  The indices $k,k^\prime$ and $a,b$ run from 1 to 4
and from 1 to 6 respectively. The Feynman rules for the counterterm
diagrams (last diagram) can be found in Eqs.~(5.1) and (5.2) 
of \rf{cscp}.}\label{feyn:diag}
\end{center}
\end{figure}

The relevant  box and penguin diagrams are depicted in
\fig{feyn:diag}, where  $H^0,h^0,A^0$ and $G^0$ are the 
neutral Higgs and would-be-Goldstone bosons respectively, $\sel_a$ are
the  charged sleptons, $\sdown_a$ denote 
the down-type squarks, $\neu_k$ are the neutralinos, and $\glu$
represents  the gluino. We perform the 
calculation in the 't Hooft-Feynman gauge, using the Feynman rules of
\rf{feyn:rules}, and adopting  the on-shell renormalization
prescription  described in \rf{cscp}.

In our subsequent calculation, we will exploit the tree-level relations 
\be\label{relation:tree-level}
 M_{A^0}^2=M_H^2-M_W^2, \quad M_{H^0}^2+M_{h^0}^2=M_{A^0}^2+M_Z^2,
\ee
\be
 \frac{\sin 2\a}{\sin 2\b} = -\left(\frac{M^2_{H^0}+M^2_{h^0}}{M^2_{H^0}-M^2_{h^0}}\right), \quad 
 \frac{\cos 2\a}{\cos 2\b} = -\left(\frac{M^2_{A^0}-M^2_Z}{M^2_{H^0}-M^2_{h^0}}
\right),
\ee
where $M_{A^0}$ and $M_H$ are the masses of the $\cp$-odd and charged
Higgs  boson respectively. $M_{h^0,H^0}$ and $\a$ are the masses and
mixing angle in the $\cp$-even Higgs sector. This leaves two 
free parameters in the Higgs sector that we choose to be  $M_{H}$ and $\tan\b$.

Introducing, for convenience, the dimensionless variables 
\be
 x_a=\frac{m_{\sdown_a}^2}{M_{\glu}^2},\quad
 x_{ak}=\frac{m_{\sdown_a}^2}{M_{\neu_k}^2},
\ee 
our results for the gluino and neutralino contributions at large $\tan\b$ 
can be summarized as follows.\footnote{Our results for the gluino and neutralino
contributions differ somewhat from those given in
Ref. \cite{xiong:yang}, which contains typographical errors 
in the formulae for the various Wilson coefficients 
\cite{privateXiong}.}

(i) Gluino:
\bea\label{wil:glu:CsCp}
 c_{S,P}^{\glu} &=&\pm
 \f{1}{V_{tb}^{}V_{tq}^{*}}\f{4\a_s}{3\a}
 \f{m_l \tan^2\b}{m_b (M_{H}^2-M_W^2)M_{\glu}}
 \sum_{a,b=1}^{6}(\G^{D_L\dag})_{qb}(\G^{D_R})_{a3}\nnu\\      
 &\times& \Bigg\{  D_2  ( x_a, x_b )\left[\G^{D_L}M_DA_D^*\G^{D_R\dag} \pm \hc\right]_{ba}-
   M_{\glu}^2 D_3(x_a)\d_{ab}]\Bigg\},
\eea
\bea\label{wil:glu:CsCp:Prime}
 c_{S,P}^{\prime\glu} &=&\pm
 \f{1}{V_{tb}^{}V_{tq}^{*}}\f{4\a_s}{3\a}
 \f{m_l\tan^2\b }{m_q (M_{H}^2-M_W^2)M_{\glu}}
 \sum_{a,b=1}^{6}(\G^{D_R\dag})_{qb}(\G^{D_L})_{a3}\nnu\\  
 &\times&  \Bigg\{D_2  ( x_a, x_b )\left[\G^{D_L}M_DA_D^*\G^{D_R\dag} \pm \hc  \right]_{ba}\mp 
 M_{\glu}^2 D_3(x_a)\d_{ab}\Bigg\},
\eea
where the  subscript $q$ on the $\G$ matrices (see Appendix
\ref{def:mass:matrices}  for details) is equal 
to $1$ $(2)$ for $\B_d$ $(\B_s)$ decays. Furthermore, $\theta_W$ is
the Weinberg angle and the functions $D_{2,3} $ are given in Appendix
\ref{aux:funcs}.  Observe that in Eqs.~(\ref{wil:glu:CsCp}) 
and (\ref{wil:glu:CsCp:Prime}) we have retained the leading and
subleading  terms in $\tan\b$, which  
might become comparable in size in some part of the MSSM parameter
space. A term is called leading or subleading if it has the following structure:
\be
C_i {\cal O}_i|_{\rm leading} \sim \tan^{n+1}\b \;
  \left(\frac{m}{M}\right)^n,\qquad
C_i {\cal O}_i|_{\rm subleading} \sim \tan^{n}\b \;
  \left(\frac{m}{M}\right)^n,
\label{lead:sublead}
\ee
$n=0,1,2$,\ldots. Further, $m$ denotes lepton and light quark masses while
$M$ stands for masses of particles that have been integrated out.

(ii) Neutralino:
\bea\label{wil:neu:CsCp}
 c_{S,P}^{\neu} &=&  \pm \f{1}{V_{tb}^{}V_{tq}^{*}}
     \f{m_l}{12M_W(M_{H}^2-M_W^2)\sin^2\theta_W}
     \sum_{k=1}^{4}\sum_{a=1}^{6} M_{\neu_k}
     D_3 ( x_{ak} )\nnu\\  
 &\times&\Bigg\{\tan^4\b\f{3 m_q}{M_W} N_{k3}^2 (\G^{D_R\dag})_{qa}(\G^{D_L})_{a3}\nnu\\
 &&+
 \tan^3\b   N_{k3}  \Bigg[( \tan\theta_W N_{k1}-3 N_{k2} ) (\G^{D_L\dag})_{qa}(\G^{D_L})_{a3}
 +\f{2 m_q}{m_b}\tan\theta_W N_{k1} 
 (\G^{D_R\dag})_{qa}(\G^{D_R})_{a3}  \Bigg]\nnu\\
 &&+\tan^2\b\f{2M_W}{3 m_b} \tan\theta_W N_{k1} \left(\tan\theta_W N_{k1}-3
   N_{k2}\right) (\G^{D_L\dag})_{qa}(\G^{D_R})_{a3}\Bigg\},
\eea
\bea\label{wil:neu:CsCp:Prime}
 c_{S,P}^{\prime\neu} &=& \f{1}{V_{tb}^{}V_{tq}^{*}}
     \f{m_l}{12M_W(M_{H}^2-M_W^2)\sin^2\theta_W}
     \sum_{k=1}^{4}\sum_{a=1}^{6} \Bigg\{ M_{\neu_k}
     D_3 (x_{ak})\nnu\\
 &\times&\Bigg\{\tan^4\b\f{3m_b}{M_W} N_{k3}^{*2} (\G^{D_L\dag})_{qa}(\G^{D_R})_{a3}\nnu\\
 &&+
 \tan^3\b   N^*_{k3}  \Bigg[( \tan\theta_W N^*_{k1}-3 N^*_{k2} ) (\G^{D_L\dag})_{qa}(\G^{D_L})_{a3}
 +\f{2 m_b}{m_q}\tan\theta_W N^*_{k1} 
 (\G^{D_R\dag})_{qa}(\G^{D_R})_{a3}  \Bigg]\nnu\\
 &&+\tan^2\b\f{2M_W}{3 m_q} \tan\theta_W N^*_{k1} \left(\tan\theta_W N^*_{k1}-3
   N^*_{k2}\right) (\G^{D_R\dag})_{qa}(\G^{D_L})_{a3}\Bigg\},
\eea
where $N$ is the neutralino mixing matrix, defined in Appendix
\ref{def:mass:matrices}. Unlike the gluino 
contributions, we have kept only the leading term in $\tan\b$
[cf. Eq. (\ref{lead:sublead})], which stems from the counterterm of the 
electroweak wave function renormalization (see
\fig{feyn:diag}). However, we have checked  numerically whether the
subleading term is important. 

\newpage

\section{Numerical analysis}\label{num:analysis}
\subsection {Experimental constraints}\label{exp:constraints}
Currently available data on rare $B$ decays already constrain
new-physics  contributions to the various Wilson coefficients
governing the  $b\to s$ transitions.\footnote{A recent analysis of
exclusive  and inclusive rare $B$ decays within SUSY has been
performed in  \rf{enrico:rare:Bdecays} by using the SM operator basis. 
Recall that in the present work we utilize an extended operator basis.}
We take  into account the following 
experimental upper limits \cite{exp:incl:excl,exp:bkmumu}:
\be\label{exp:BXsee}
 \br(\B\to X_s e^+ e^-) < 10.1\times 10^{-6} \quad (\cl{90}),
\ee
\be
 \br(\B\to X_s \m^+\m^-) <  19.1 \times 10^{-6} \quad (\cl{90}),
\ee
\be
 \br(\B\to K^{*} \m^+\m^-) < 3.1 \times 10^{-6} \quad (\cl{90}),
\ee
besides those on $\Bds$ decays given in
Eqs.~(\ref{Bmumu:exp:bounds}). As far  as the measured branching fractions 
of $\B\to K \m^+\m^-$ \cite{exp:bkmumu} and $\B\to X_s\gamma$
\cite{exp:bsg}  are concerned, we will allow the following ranges:
\be  
 0.5 \times 10^{-6} \leqslant  \br(\B\to K \m^+\m^-) \leqslant  1.5 
\times 10^{-6},
\ee
\be\label{exp:bsg}
 2.0\times 10^{-4} \leqslant  \br(\B\to X_s \g) \leqslant 4.5 \times 10^{-4}.
\ee
The constraints on the flavour-changing entries in the matrices
$\ml{U}$ and  $\ml{D}$ will be discussed below.
\subsection{Implications of \boldmath${{\overline{\rm MFV}}}$ for the decays
\bm $\B_{\lowercase{d,s}}\to \m^+\m^-$}\label{sec:scenario:B}
In our analysis, we will perform a scan over the following ranges of
MSSM parameters:
\begin{mathletters}\label{scan:range}
\be
 125\ \GeV \leqslant M_H \leqslant 500\ \GeV,
\ee
\be
 100\ \GeV \leqslant M_{1,2} \leqslant 1\ \TeV,\quad 250\ \GeV
 \leqslant  M_{\glu} \leqslant 1\ \TeV,
\ee
\be
 -1\ \TeV \leqslant \left\{ \mu,\,(A_U)_{ii},\,(A_D)_{ii} \right\} 
 \leqslant 1\ \TeV, 
\ee
\be
 300\ \GeV \leqslant \left\{ \mlii{U},\,\mlii{D},\,
   \mrii{U},\,\mrii{D} \right\} \leqslant 1\ \TeV.
\ee\end{mathletters}%
Further, we fix $\tan\b=50$ and $m_{\tilde l, \tilde \n}=100\ \GeV$,
and take the lower bounds on the sparticle masses from \cite{pdg}. 
As far as the decay constants $f_{B_q}$ are concerned, we will take
the central values as given in \eq{decay-constants}. The remaining
input parameters are fixed as given in Tables 1 and 2 
of \rf{nlo:susy:RareDecays}. 

We have analyzed more than 100,000 parameter sets, which have been
produced randomly. The so-called unconstrained plots contain those
parameter sets that are consistent with the lower bounds on the
sparticle masses. On the other hand, in the constrained plots we have
also taken into account the bounds from \eqs{exp:BXsee}{exp:bsg} as
well as the constraints on the flavour-changing entries in the matrices
$\ml{U}$ and  $\ml{D}$ (see discussion below).

As for the various branching ratios, it turns out that the primed Wilson
coefficients are negligibly small, and hence can be safely neglected.
[Comments on this issue can be found at the end of the discussion of
scenario (C).]

\subsection*{Scenario (A)}
\begin{figure}[thb]
\begin{center}
\mbox{\epsfxsize=6.5cm\epsffile{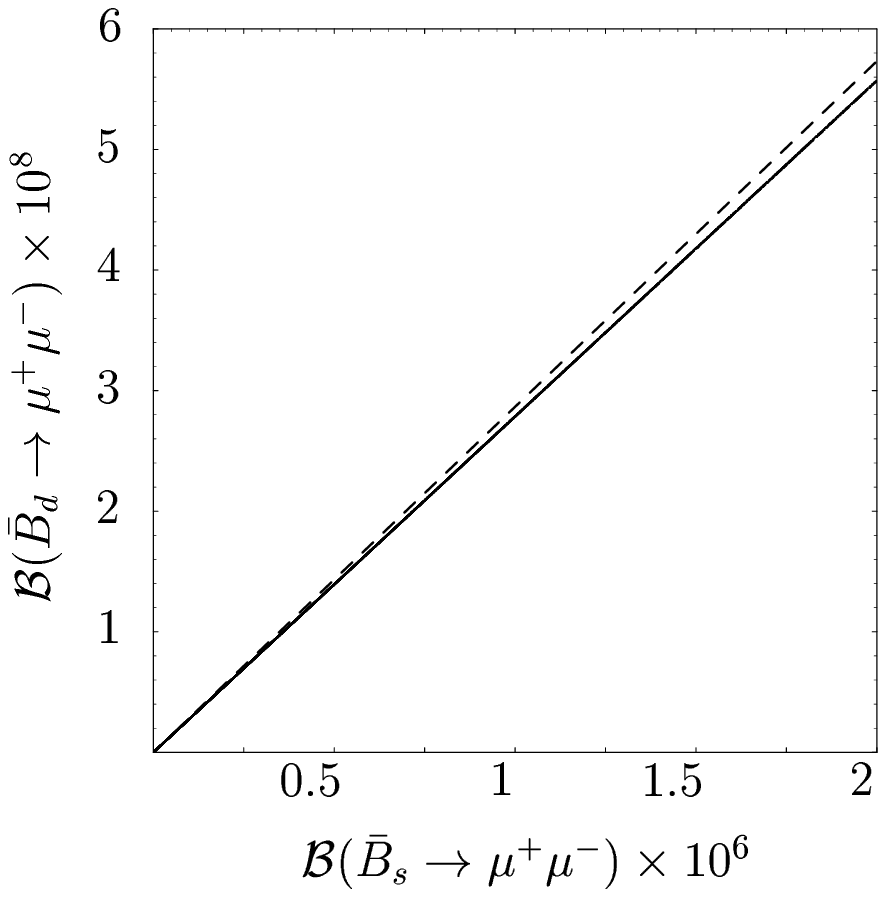}}
\vspace{3mm}
\caption{Predictions for the branching ratios $\br(\Bd)$
  vs. $\br(\Bs)$ in scenario (A). The dashed line corresponds to $R_{\rm
    SM}\approx 0.03$  while the solid line is the result for
  this scenario. The ratio $R\equiv\br(\Bd)/\br(\Bs)$ varies
  between $0.028\lesssim R\lesssim 0.029$. Since the results for the
  constrained and unconstrained
  plots are indistinguishable, we only show the former.
\label{scatterscenA}}
\end{center}
\end{figure}

Recall that in scenario (A) the matrices $\ml{U}$ and $\ml{D}$ are
equal  and proportional to the unit matrix. Therefore, the gluinos and
neutralinos  do not contribute at one-loop level. In
Fig. \ref{scatterscenA}, we have plotted  $\br(\Bd)$ versus
$\br(\Bs)$, where the dashed line represents our 
reference curve with the slope $R_{\rm SM}$ while the solid line is
our  result obtained within scenario (A). The scan over the 
parameter region given in Eqs. (\ref{scan:range}) shows that the ratio $R$
is approximately constant and close to $R_{\rm SM}\approx
0.03$.  

\subsection*{Scenario (B)}
In this scenario, the matrix $\ml{D}$ is diagonal,
\be
 \ml{D} = \diag(m^2_{\tilde d_{L}},m^2_{\tilde s_{L}},m^2_{\tilde b_{L}}),
\ee
with at least two different entries; hence there are no gluino and
neutralino contributions. Employing the relation in
\eq{su2}, the  matrix $\ml{U}$ becomes
\be\label{mull:b}
\ml{U} = \vckm  \ml{D} \vckm^\dag = \left(\begin{array}{ccc} 
         m^2_{\tilde u_{L}} & \delu{12} & \delu{13} \\ 
         \delu{12*} & m^2_{\tilde c_{L}} & \delu{23} \\ 
         \delu{13*} & \delu{23*} & m^2_{\tilde t_{L}}\end{array}\right),
\ee 
with the off-diagonal elements 
\begin{mathletters}\label{fcnc:entries:up}
\be
\delu{12} = (m^2_{\tilde d_{L}}-m^2_{\tilde s_{L}})V_{ud}^{}V_{cd}^*+ 
            (m^2_{\tilde b_{L}}-m^2_{\tilde s_{L}})V_{ub}^{}V_{cb}^*,
\ee
\be
\delu{13} = (m^2_{\tilde d_{L}}-m^2_{\tilde s_{L}})V_{ud}^{}V_{td}^*+ 
            (m^2_{\tilde b_{L}}-m^2_{\tilde s_{L}})V_{ub}^{}V_{tb}^*,
\ee
\be
\delu{23} = (m^2_{\tilde d_{L}}-m^2_{\tilde s_{L}})V_{cd}^{}V_{td}^*+ 
            (m^2_{\tilde b_{L}}-m^2_{\tilde s_{L}})V_{cb}^{}V_{tb}^*.
\ee
\end{mathletters}%
In writing these equations, we have used the unitarity of the CKM
matrix.  The flavour-changing entries given above, together with the
corresponding elements in the down squark sector [see Eqs.~(\ref{fcnc:entries})
below], are constrained by experimental data on $K^0$--$\bar K^0$, $
B^0$--$\B^0$,  $D^0$--$\bar D^0$  oscillations, and the $b\to s \g$
decay \cite{fcnc,bsg:deltas:allcontribs,deltas:bounds}.\footnote{Strictly
  speaking, these constraints are only valid if all squark masses are
  close in size,  but the order of magnitude should also be valid 
  for non-degenerate masses \cite{communication}.}

\begin{figure}[thb]
\begin{center}
\begin{tabular}{cc}
\mbox{\epsfxsize=6.5cm\epsffile{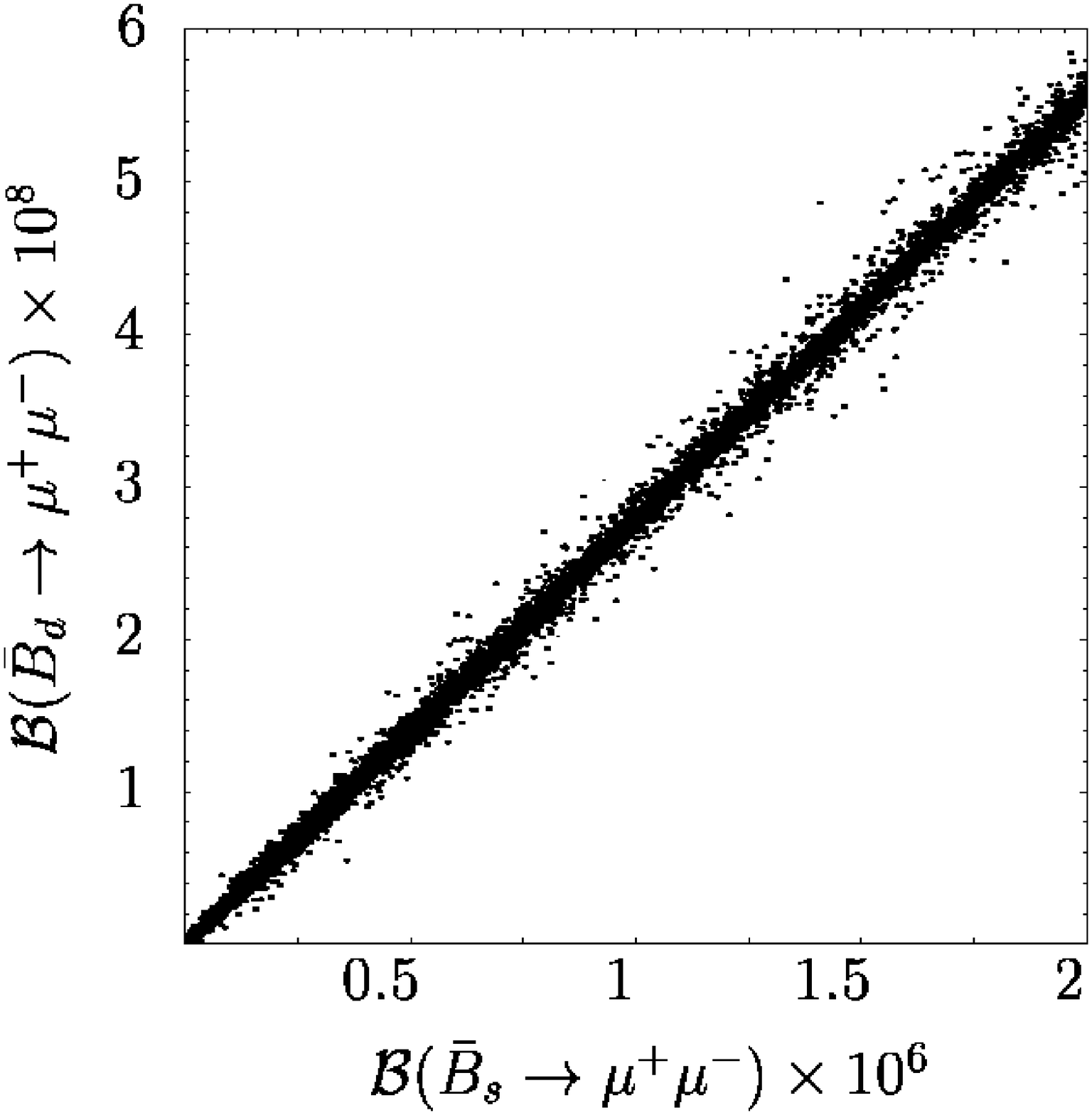}}\hspace{1cm}&
\hspace{1cm}
\mbox{\epsfxsize=6.5cm\epsffile{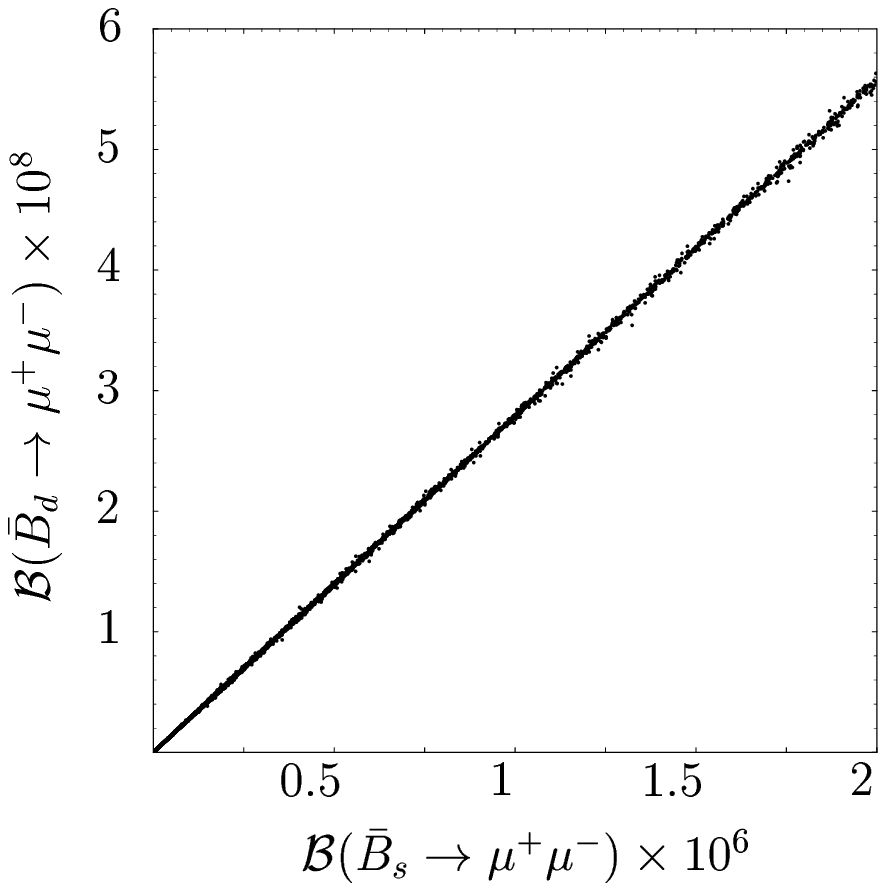}}
\end{tabular}
\vspace{3mm}
\caption{Predictions for the branching ratios $\br(\Bd)$
  vs. $\br(\Bs)$ in scenario (B). The left plot shows the
  unconstrained parameter sets, with $R$ varying between
  $0.002\lesssim R\lesssim 0.115$. The right plot exhibits the results
  for the branching fractions, using the constrained parameter sets. In
  this case, $R$ varies between $0.026$ and $0.030$.}
\label{scatterscenB}
\end{center}
\end{figure}

In Fig.~\ref{scatterscenB}, we have plotted $\br(\Bd)$ versus 
$\br(\Bs)$ for unconstrained and constrained parameter sets. 
In case of the former, the deviation of $R$ from $R_{\rm SM}$ 
can be one order of magnitude, with the range
$0.002\lesssim R\lesssim 0.115$. Applying the constraints 
on the flavour-changing entries in the matrix $\ml{U}$, as well as 
the bounds given in \eqs{exp:BXsee}{exp:bsg}, our numerical predictions
for $R$ in scenario (B) are similar to those in scenario (A). In fact,
$R$ varies between $0.026$ and $0.030$.
It is important to note that the bounds on $\delu{ij}$ \cite{fcnc}
severely constrain the additional chargino contributions in scenario
(B) in contrast to scenario (A). 

\subsection{Scenario (C)}
We now repeat the analysis of the $\Bds$ decays within the framework
of scenario (C). 
In this case, the matrix $\ml{U}$ is diagonal,
\be
 \ml{U} = \diag(m^2_{\tilde u_{L}},m^2_{\tilde c_{L}},m^2_{\tilde t_{L}}),
\ee 
with at least two different entries. According to the relation in
\eq{su2},  this implies that $\ml{D}$ has non-diagonal entries, so
that  gluinos and neutralinos contribute to the $b\to ql^+l^-$
transition already at one-loop level. In this case, $\ml{D}$ can be written as
\be\label{mdll:b}
\ml{D} = \vckm^\dag \ml{U} \vckm = \left(\begin{array}{ccc} 
         m^2_{\tilde d_{L}} & \del{12} & \del{13} \\ 
         \del{12*} & m^2_{\tilde s_{L}} & \del{23} \\ 
         \del{13*} & \del{23*} & m^2_{\tilde b_{L}}\end{array}\right),
\ee 
where the flavour-changing off-diagonal entries are given by 
\begin{mathletters}\label{fcnc:entries}
\be
\del{12} = (m^2_{\tilde u_{L}}-m^2_{\tilde c_{L}})V_{us}^{}V_{ud}^*+ 
                (m^2_{\tilde t_{L}}-m^2_{\tilde c_{L}})V_{ts}^{}V_{td}^*,
\ee
\be
\del{13} = (m^2_{\tilde u_{L}}-m^2_{\tilde c_{L}})V_{ub}^{}V_{ud}^*+ 
                (m^2_{\tilde t_{L}}-m^2_{\tilde c_{L}})V_{tb}^{}V_{td}^*,
\ee
\be
\del{23} = (m^2_{\tilde u_{L}}-m^2_{\tilde c_{L}})V_{ub}^{}V_{us}^*+ 
                (m^2_{\tilde t_{L}}-m^2_{\tilde c_{L}})V_{tb}^{}V_{ts}^*.
\ee
\end{mathletters}%
As before, we take the constraints of 
\rfs{fcnc,bsg:deltas:allcontribs,deltas:bounds} on these off-diagonal 
elements.

\begin{figure}[thb]
\begin{center}
\begin{tabular}{cc}
\mbox{\epsfxsize=6.5cm\epsffile{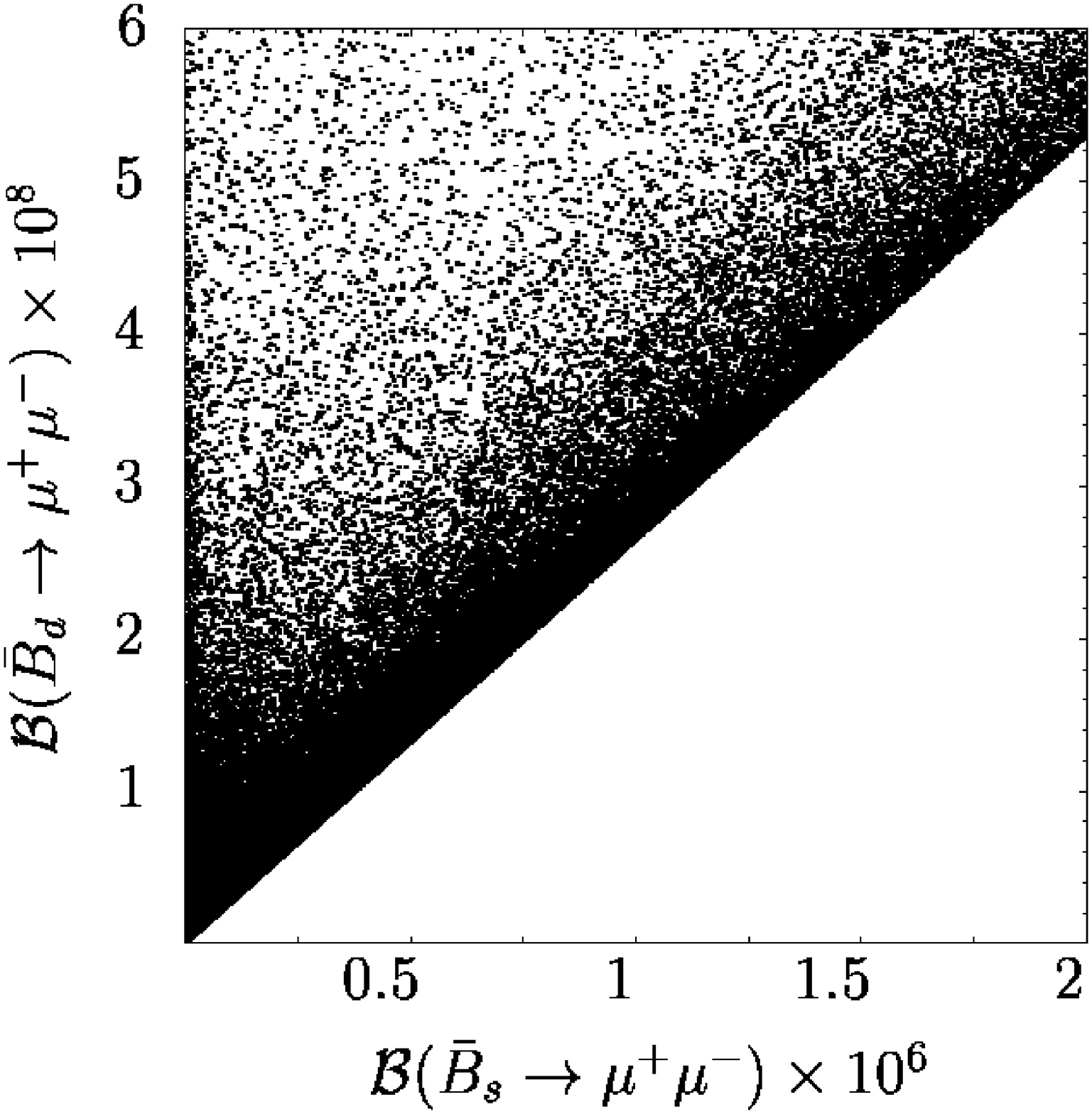}}\hspace{1cm}&
\hspace{1cm}
\mbox{\epsfxsize=6.5cm\epsffile{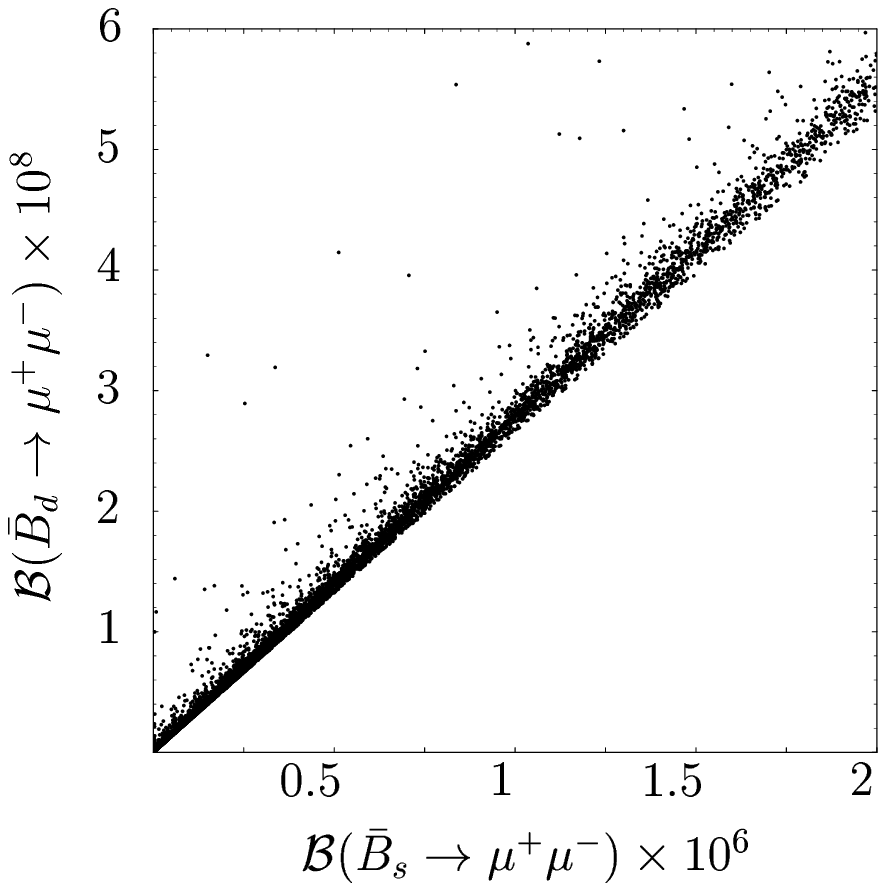}}\\[3mm]
\mbox{\hspace{-9mm}\epsfxsize=7cm\epsffile{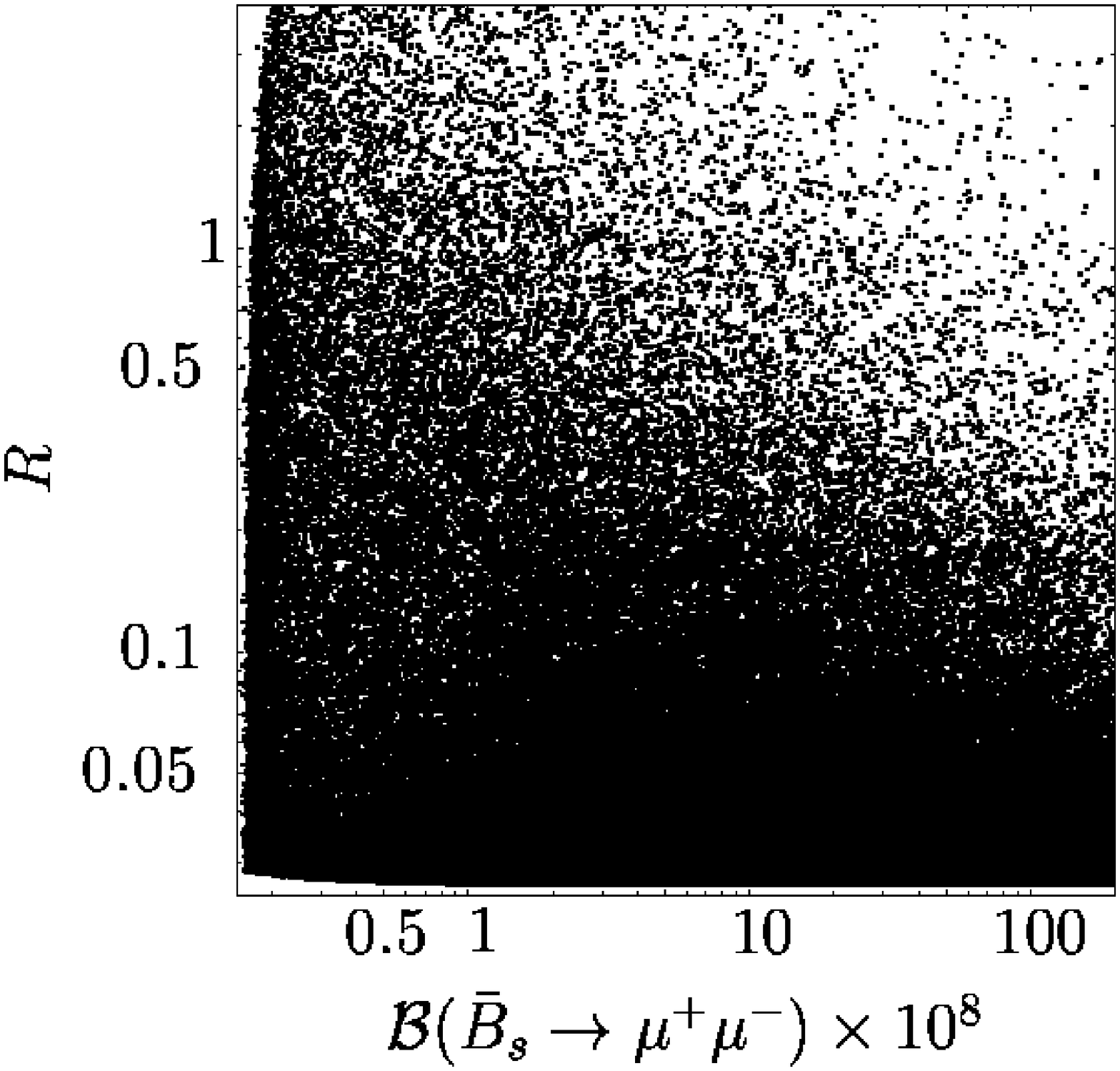}}\hspace{0.7cm}&
\hspace{0.2cm}
\mbox{\epsfxsize=7cm\epsffile{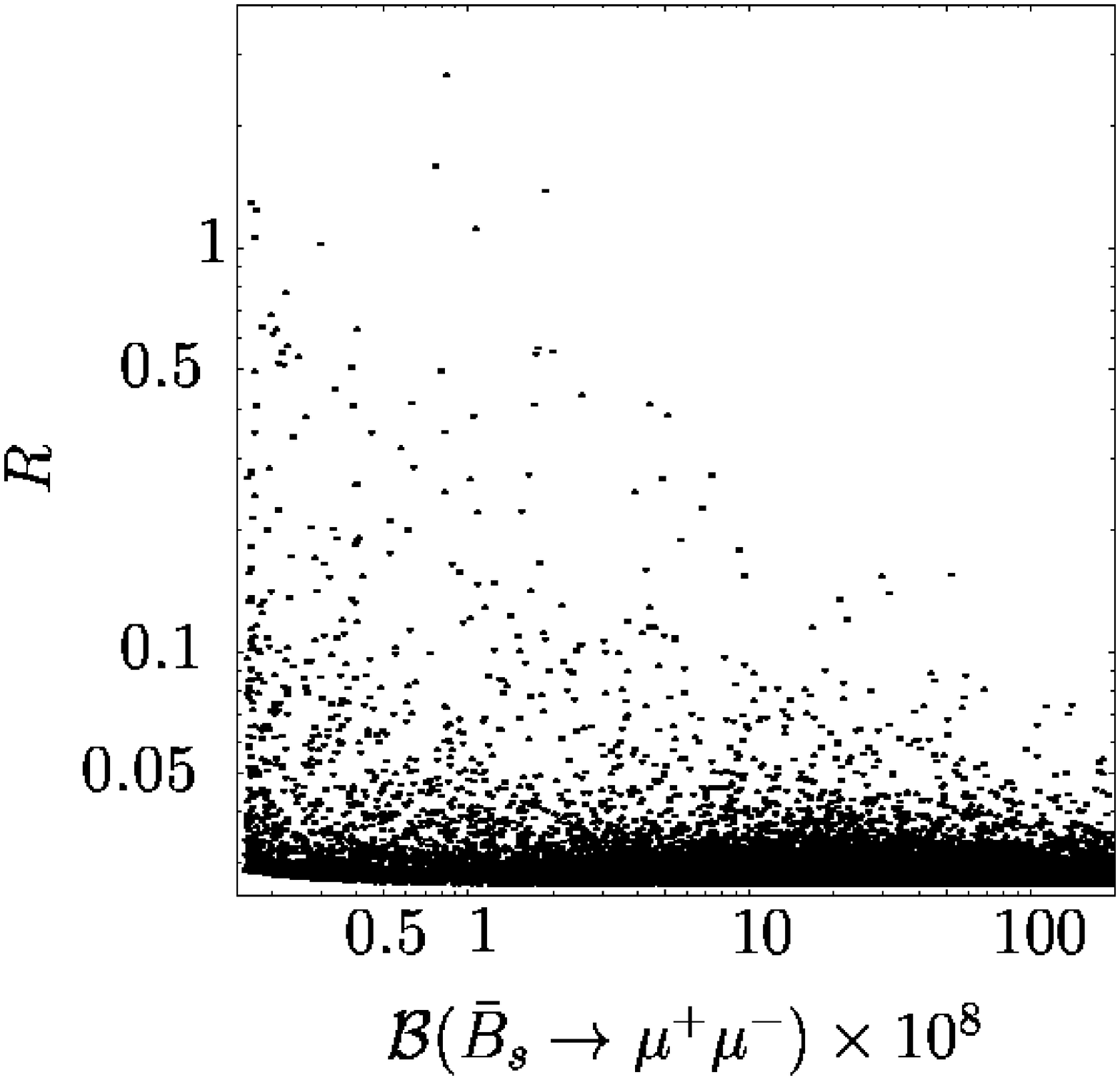}}\\
\end{tabular}
\vspace{3mm}
\caption{Predictions for the branching ratios $\br(\Bd)$
  vs. $\br(\Bs)$ (upper plots) and $R$ vs. $\br(\Bs)$ (lower plots) in 
  scenario (C). The left plots show the
  unconstrained parameter sets, with $R$ varying between
  $0.026\lesssim  R\lesssim 20.636$. The right plots exhibit the results
  for the branching fractions and their ratio $R$, using the
  constrained 
  parameter sets. In this case, $R$ varies between $0.026$ and $2.863$.}
\label{scatterscenC}
\end{center}
\end{figure}

\clearpage

The scatter plots in Fig.~\ref{scatterscenC} exhibit an order-of-magnitude
deviation from $R_{\rm SM} \approx 0.03$. In the unconstrained case
(left plots) $R$ is predicted to be in the range $0.026\lesssim
R\lesssim 20.636$, while for the constrained parameter sets (right
plots) we find $0.026\lesssim R\lesssim 2.863$. A noticeable feature of
scenario (C) is that there exists a lower bound on $R$, i.e. 
$R\gtrsim 0.95\,R_{\rm SM}$ (see upper plots of
Fig.~\ref{scatterscenC}), which is due to the structure of the CKM 
matrix. We stress that
this bound is valid only within scenario (C) and does not apply to 
scenario (B) or scenarios with new sources of flavour violation
(see Sec. \ref{mssm}). 

\begin{figure}[thb]
\begin{center}
\begin{tabular}{cc}
\mbox{\epsfxsize=6.5cm\epsffile{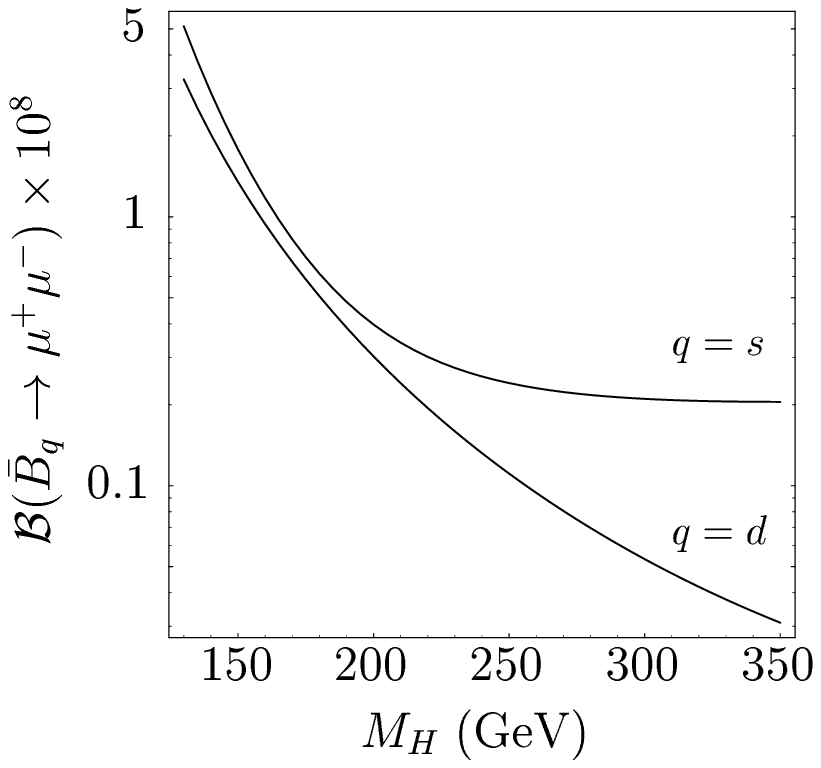}}\hspace{1cm}&
\hspace{1cm}
\mbox{\epsfxsize=6.5cm\epsffile{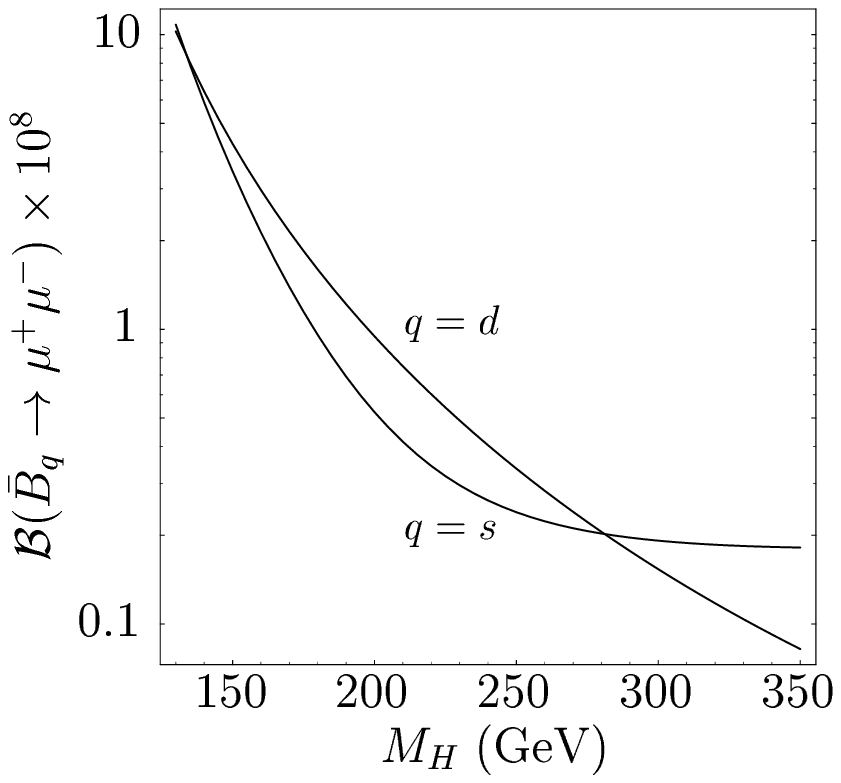}}
\end{tabular}
\vspace{3mm}
\caption{$\br(\Bds)$ as function of the charged Higgs boson mass,
  $M_H$, for $\tan\b=50$
 (left plot) and $\tan\b=60$ (right plot), taking into account the 
  experimental constraints on the rare $B$ decays and on the flavour-changing 
  entries. The remaining parameters have been kept fixed, as described 
  in the text.}
\label{mfvc:Bq-mh}
\end{center}
\end{figure}

In \fig{mfvc:Bq-mh}, the branching ratios of $\Bd$ and $\Bs$ are displayed as 
functions of the charged Higgs boson mass, $M_H$, for $\tan\b$ equal to $50$ 
(left plot) and $60$ (right plot). The remaining parameters are given by
\begin{mathletters}
\be
 \m = -800\ \GeV, \ M_1 = 500\ \GeV,\ M_2 = 200\ \GeV, \ M_{\glu} = 250\ \GeV,
\ee
\be
 A_U = A_D = {\rm diag} ( 100\ \GeV,\ 100\ \GeV,\ 75\ \GeV),
\ee
\be
 M_{\tilde{U}_R} = M_{\tilde{D}_R} = {\rm diag} ( 500\ \GeV,\ 500\
 \GeV,\ 500\ \GeV),
\ee 
\be
 M_{\tilde{U}_L} = {\rm diag} ( 460\ \GeV,\ 500\ \GeV,\ 517\ \GeV).
\ee\end{mathletters}%
Both plots are consistent with the constraints on rare $B$ decays 
in \eqs{exp:BXsee}{exp:bsg}, and with those on the flavour-changing 
entries $\del{ij}$ \cite{fcnc}. Interestingly, in the left (right) 
plot, the ratio $R$ ranges between $0.15 \lesssim R \lesssim 0.81$ 
($0.44 \lesssim R \lesssim 1.78$), while the magnitude of the individual
branching fractions decreases drastically with increasing charged Higgs 
boson mass, $M_H$. Note that for small $M_H$ and
$\tan\b$ close to $60$ both branching ratios are in a region that
can be probed experimentally, in Run II of the Fermilab Tevatron,
BABAR and Belle.
The large enhancement of $R$ is due to
$\Im[c_{S,P}^{\glu}(\Bd)]$ being by an order-of-magnitude larger than 
$\Im[c_{S,P}^{\glu}(\Bs)]$ and a
cancellation between the real parts of  $c_{S,P}^{\chargino}$ and
$c_{S,P}^{\glu}$. As a consequence, the neutralino contributions
to the scalar and pseudoscalar coefficients, although negligible
compared to
the corresponding contributions in the chargino and gluino sector,
cannot be neglected in the $\Bs$ decay. The neutralinos contribute
significantly to the ratio $R$; for illustration, the ratio
$R\approx0.94$ in the right plot ($\tan\b = 60$) at $M_H=130\ \GeV$
increases to $R\approx1.55$ after neglecting the contributions from
neutralinos.

As already mentioned, we have found that the primed Wilson
coefficients, $c_{S,P}^{\prime}$, are numerically
suppressed. In fact, applying the mass insertion approximation as outlined in
Appendix \ref{app:pert:diag}, one can convince oneself that
the enhancement factor $1/m_q$ (recall $q=s,d$) appearing in the
scalar and pseudoscalar coefficients cancels out. Therefore, the primed
Wilson coefficients are  suppressed by a factor
$m_q/m_b\ll 1$, compared to the unprimed coefficients.

\section{Summary and conclusions}\label{discussion}
In the framework of the minimal supersymmetric extension of the SM
with large $\tan\b$, 
we have computed  the gluino and neutralino exchange diagrams
contributing to  the purely leptonic decays $\Bds$. Together with our
previous analysis \cite{cscp}, the present paper provides a complete
study of the decays $\Bds$ at one-loop level in the MSSM with
modified minimal flavour violation  and large $\tan\b$. 

We have defined $\overline{\rm MFV}$  using symmetry
arguments and have shown that $\overline{\rm MFV}$ is less 
restrictive than MFV, while 
the CKM matrix remains the only source of flavour violation. We have given 
a criterion for testing whether a theory belongs to the class of 
$\overline{\rm MFV}$.
Within the MSSM we have investigated three scenarios that are possible 
within the context of $\overline{\rm MFV}$.  
In particular, we have studied the
case  where the gluino and neutralino exchange diagrams contribute besides
$W^\pm,H^ \pm,\chargino$ [scenario (C)]. The neutralino
Wilson coefficients are numerically smaller than those coming from the
chargino and gluino contributions. However, we have found that in
certain  regions of the MSSM parameter space cancellations
between the chargino and gluino coefficients occur, in which case the
neutralino contributions become important. As a matter of fact, for the
SUSY parameter sets examined, we found that a large
value of $R\equiv\br(\Bd)/\br(\Bs)$ always involves such a
cancellation.  

Including current experimental data on rare $B$ decays, as well as on
$K, B, D$ meson  mixing, we found that in certain regions of the SUSY
parameter space the branching ratios $\br(\Bd)$ and $\br(\Bs)$ can be
up to the order of $10^{-7}$ and $10^{-6}$ respectively. 
Specifically, we showed that there exist regions in
which the branching fractions of both decay modes are comparable in
size, and may well be accessible to Run II of the Fermilab Tevatron.  

We wish to stress that a measurement of the branching ratios $\br(\Bds)$, 
or equivalently,  a ratio $R$ of $O(1)$, does not necessarily imply
the  existence of new flavour violation outside the CKM matrix.
Nevertheless, any observation of these decay modes in ongoing and forthcoming 
experiments would be an unambiguous signal of new physics.

\acknowledgments
We would like to thank Andrzej J.~Buras, Manuel~Drees, Gino Isidori, and Janusz Rosiek
for  useful discussions.~We are grateful to Andrzej J.~Buras for his
comments  on the manuscript. We would also like to thank Martin Gorbahn for suggesting
the name {\em Modified Minimal Flavour Violation}.
This work was supported in part by the 
German `Bundesministerium f\"ur Bildung und Forschung' under contract 
05HT1WOA3 and by the `Deutsche Forschungsgemeinschaft' (DFG) under 
contract Bu.706/1-1.

\begin{appendix}

\section{Auxiliary functions}\label{aux:funcs}
The loop functions appearing in the formulae of \sec{computation} are given by 
\cite{cscp}
\be
 D_2(x,y)= \frac{x\ln x}{(1-x)(x-y)}+(x\leftrightarrow y),
\ee
\be
 D_3(x)= \frac{x\ln x}{1-x}.
\ee
\section{Mass matrices}\label{def:mass:matrices} 
\subsection{Squark mass matrices}
The squark mass-squared matrices given in \eq{squark:mass} are
diagonalized according to
\be
 \G^{U} \M_U^2 \G^{U\dag} = 
   \diag(m_{\sup_1}^2,\dots,m_{\sup_6}^2),\quad
 \G^{D} \M_D^2 \G^{D\dag} = 
   \diag(m_{\sdown_1}^2,\dots,m_{\sdown_6}^2). 
\ee
It is convenient to split the $6 \times 6$ squark mixing matrices, 
$\G^{U,D}$, into two $6 \times 3$ submatrices:
\be
(\G^Q)_{ai}=(\G^{Q_L})_{ai}, \quad   (\G^Q)_{a,
  i+3}=(\G^{Q_R})_{ai},\quad Q=U,D,
\ee
where  $a=1,\dots, 6$ and $ i=1,2,3$.
\subsection{Neutralino mass matrix}
The neutralino mass matrix has the structure \cite{mssm}   
\be
 \M_{\neu} = \left( \begin{array}{cccc} 
     M_1 & 0 & -M_Z \sin\theta_W \cos\b & M_Z \sin\theta_W \sin\b \\
     0 & M_2 & M_Z \cos\theta_W \cos\b & -M_Z \cos\theta_W \sin\b \\
     -M_Z \sin\theta_W \cos\b & M_Z \cos\theta_W \cos\b & 0 & -\m \\
     M_Z \sin\theta_W \sin\b & -M_Z \cos\theta_W \sin\b & -\m & 0 
 \end{array} \right),
\ee
which is diagonalized by a unitary matrix $N$ such that
\be\label{def:N}
 N^{\ast} \M_{\neu} N^{\dag} = 
  \diag(M_{\neu_1}, M_{\neu_2}, M_{\neu_3}, M_{\neu_4}).
\ee

\section{Perturbative diagonalization of \bm $\M_D^2$}\label{app:pert:diag}
Within the framework of scenario (C), all flavour-changing interactions and 
CP violation are due exclusively to the CKM matrix.~In this case, the
matrices  $\M^2_{D_{LR}}$ and $\M^{2}_{D_{RR}}$ in \eq{squark:mass}
are  diagonal and real whereas $\M^2_{D_{LL}}$ contains complex off-diagonal  
entries [cf.~\eq{mdll:b}]. In order to  diagonalize the down squark  
mass-squared matrix $\M^2_{D}$ perturbatively, we rewrite it as  
\be\label{mass:matrix:down:sector}
 \M_D^2 = \ti{\M}_D^2 + \ti{\D}_D =
 \Bigg( \begin{array}{cc} \ti{\M}^2_{D_{LL}} & \M^2_{D_{LR}} \\ 
                \M^{2}_{D_{LR}} & \M^2_{D_{RR}}  
        \end{array} \Bigg)+
 \Bigg( \begin{array}{cc} \D_{D_{LL}} & 0 \\ 0 & 0 \end{array} \Bigg),
\ee 
where $\ti{\M}^2_{D_{LL}}$ (${\D}_{D_{LL}}$) contains the diagonal
(off-diagonal) elements of $\M^2_{D_{LL}}$. We remark parenthetically
that  for large values of $\tan\b$ the LR elements in $\M^2_{D}$
cannot  in general be treated as small perturbations. [This is the
reason for choosing the decomposition of $\M^2_{D}$ given in  
\eq{mass:matrix:down:sector}.]

Writing $\G^D$ as a product of two unitary matrices, $\G^D\equiv XY$, we have
\be\label{pert:diag}
 (\M_D^2)_\diag = \G^{D}\M_D^2\G^{D\dag} 
 = X \left\{ D+\D_D \right\} X^\dag,
\ee 
where $D=Y\ti{\M}_D^2Y^\dag\equiv \diag(D_1,\dots,D_6)$ and  
$\D_D=Y\ti{\D}_DY^\dag$. Then, if $|(\D_D)_{ij}| \ll D_a$, we can make
the ansatz 
\bea
 X &=& \openone +\d X^{(1)}+\d X^{(2)}+\cdots, \nnu \\
 (\M_D^2)_\diag &=& D+\d D^{(1)}+\d D^{(2)}+\cdots
\eea
to solve \eq{pert:diag} perturbatively in terms of $\D_D$. As a result, we obtain
\bea\label{mia:exp}
 \G_{ac}^{D\dag}f(m^2_{\sdown_c})\G^D_{cb}
 &=& Y_{ac}^\dag f(D_c) Y_{cb}^{}+Y_{ac}^\dag \D_{cd} f_1(D_c,D_d) Y_{db}^{}\nnu\\
 &+&Y_{ac}^\dag \D_{ce}\D_{ed}f_2(D_c,D_d,D_e) Y_{db}^{}+\cdots,
\eea
where $f$ denotes an arbitrary loop function, and
\bea
 f_1(x,y) = \f{f(x)-f(y)}{x-y} ,\quad 
 f_2(x,y,z) = \f{f_1(x,z)-f_1(y,z)}{x-y},
\eea
for $x\neq y \wedge x\neq z \wedge y\neq z$. In all other cases the corresponding limit of the functions $f_{1,2}$
has to be taken. Setting $Y \equiv \openone$ in \eq{mia:exp} reproduces the result given in \rfs{mia}.
\end{appendix}



\begin{thebibliography}{99}

\bibitem{lectures}A.~J.~Buras, in 
\emph{Flavour Dynamics: CP Violation and Rare Decays}, Lectures given at 
International School of Subnuclear Physics, Erice, Italy, 2000, hep-ph/0101336.

\bibitem{cscp}C.~Bobeth, T.~Ewerth, F.~Kr\"uger, and J.~Urban, \prd{64}{2001}{074014}. 

\bibitem{lattice}UKQCD Collaboration, K.~C.~Bowler \ea, \np{619}{2001}{507};
C.~Bernard, \npps{94}{2001}{159}; C.~T.~Sachrajda, \nim{462}{2001}{23};
S.~Ryan,  \npps{106}{2002}{86}.

\bibitem{exp:bmumu}Belle Collaboration, Report No. BELLE-CONF-0127 (unpublished); 
CDF Collaboration, F.~Abe \ea, \prd{57}{1998}{3811}.

\bibitem{Bll:SUSY}C.-S.~Huang and Q.-S.~Yan, \pl{442}{1998}{209}; 
C.-S.~Huang, W.~Liao, and Q.-S.~Yan, \prd{59}{1999}{011701};
K.~S.~Babu and C.~Kolda, \prl{84}{2000}{228};
G.~Isidori and A.~Retico,  \jhep{11}{2001}{001}.


\bibitem{chan:slaw}P.~H.~Chankowski and \L.~S\l awianowska, \prd{63}{2001}{054012}.

\bibitem{huang:etal}C.-S.~Huang, W.~Liao, Q.-S.~Yan, and S.-H.~Zhu, 
\prd{63}{2001}{114021};  \ibid{64}{2001}{059902(E)}.

\bibitem{oszi}A.~J.~Buras, P.~H.~Chankowski, J.~Rosiek and \L. S\l awianowska,
\np{619}{2001}{434}.

\bibitem{scen:B}
A.~J.~Buras, P.~H.~Chankowski, J.~Rosiek and \L. S\l awianowska,
hep-ph/0207241.

\bibitem{Bmumu:mSUGRA}A.~Dedes, H.~K.~Dreiner, and U.~Nierste, \prl{87}{2001}{251804}.

\bibitem{Bmumu:mSUGRA2}R.~Arnowitt, B.~Dutta, T.~Kamon, and M.~Tanaka, \pl{538}{2002}{121};
D.~A.~Demir, K.~A.~Olive, and M.~B.~Voloshin, hep-ph/0204119.

\bibitem{laplace}
S.~Bergmann and G. Perez, \prd{64}{2001}{115009};
A.~J.~Buras and R.~Fleischer, \prd{64}{2001}{115010};
S.~Laplace, Z.~Ligeti, Y.~Nir and G.~Perez, \prd{65}{2002}{094040};
P.~H.~Chankowski and J.~Rosiek, hep-ph/0207242.

\bibitem{uut}
A.~J.~Buras, P.~Gambino, M.~Gorbahn, S.~J\"ager and L.~Silvestrini,
\pl{500}{2001}{161}.

\bibitem{london}
A.~Ali and D.~London, \euro{9}{1999}{687}.

\bibitem{munich}
M.~Ciuchini, G.~Degrassi, P.~Gambino and G.~F.~Giudice, \np{534}{1998}{3};
A.~J.~Buras, P.~Gambino, M.~Gorbahn, S.~J\"ager and L.~Silvestrini,
\np{592}{2001}{55}.

\bibitem{def:MFV}
G.~D'Ambrosio, G.~F.~Giudice, G.~Isidori and A.~Strumia, hep-ph/0207036.

\bibitem{fcnc}M.~Misiak, S.~Pokorski, and J.~Rosiek, in 
\emph{Heavy Flavours II}, edited by A.~J.~Buras and M.~Lindner 
(World Scientific, Singapore, 1998), p.~795, hep-ph/9703442.

\bibitem{jamin:lange}M.~Jamin and B.~O.~Lange, \prd{65}{2002}{056005}.

\bibitem{feyn:rules}J.~Rosiek, \prd{41}{1990}{3464}; hep-ph/9511250.

\bibitem{xiong:yang} Z.~Xiong and J.~M.~Yang, \np{628}{2002}{193}.

\bibitem{privateXiong} Z.~Xiong (private communication).

\bibitem{enrico:rare:Bdecays}A.~Ali, E.~Lunghi, C.~Greub, and G.~Hiller, hep-ph/0112300.

\bibitem{exp:incl:excl}Belle Collaboration, K.~Abe \ea, hep-ex/0107072; 
see also BABAR Collaboration, B.~Aubert \ea, \prl{88}{2002}{241801}.

\bibitem{exp:bkmumu}Belle Collaboration, K.~Abe \ea, \prl{88}{2002}{021801}.

\bibitem{exp:bsg}ALEPH Collaboration, R.~Barate \ea, \pl{429}{1998}{169}; 
CLEO Collaboration, S.~Chen \ea, \prl{87}{2001}{251807};
Belle Collaboration, K.~Abe \ea, \pl{511}{2001}{151}.

\bibitem{pdg}Particle Data Group, D.~E.~Groom \ea, \euro{15}{2000}{1}.

\bibitem{nlo:susy:RareDecays}C.~Bobeth, A.~J.~Buras, F.~Kr\"uger, and 
J.~Urban,  \np{630}{2002}{87}.

\bibitem{bsg:deltas:allcontribs}T.~Besmer, C.~Greub, and T.~Hurth, \np{609}{2001}{359}.

\bibitem{deltas:bounds}See also F.~Gabbiani, E.~Gabrielli, A.~Masiero, and 
L.~Silvestrini, \np{477}{1996}{321}; D.~Chang, W.~F.~Chang, W.~Y.~Keung, N.~Sinha, 
and R.~Sinha, \prd{65}{2002}{055010}.

\bibitem{communication}G. Raz, hep-ph/0205310; J.~Rosiek (private communication). 


\bibitem{mssm}H.~E.~Haber and G.~L.~Kane, \prp{117}{1985}{75}.

\bibitem{mia}A.~J.~Buras, A.~Romanino, and L.~Silvestrini, \np{520}{1998}{3};
G.~Colangelo and G.~Isidori, \jhep{09}{1998}{009}.

\end{thebibliography}
\end{document}